\documentclass
[superscriptaddress,secnumarabic,amssymb,amsmath,nobibnotes,aps,prd,showkeys,showpacs,nofootinbib]{revtex4}%
\usepackage{graphicx}
\usepackage{epsf}
\usepackage{bm}
\usepackage{amsmath}
\usepackage{amsfonts}
\usepackage{amssymb}
\usepackage{epstopdf}%
\providecommand{\U}[1]{\protect\rule{.1in}{.1in}}
\newcommand{\be}{\begin{equation}}
\newcommand{\ee}{\end{equation}}

\newcommand{\mincir}{\raise
-3.truept\hbox{\rlap{\hbox{$\sim$}}\raise4.truept\hbox{$<$}\ }}
\newcommand{\magcir}{\raise
-3.truept\hbox{\rlap{\hbox{$\sim$}}\raise4.truept\hbox{$>$}\ }}

\begin{document}
\title{Conservation laws and Exact Solutions in Brans-Dicke Cosmology with a Scalar Field}
\author{Andronikos Paliathanasis}
\email{anpaliat@phys.uoa.gr}
\affiliation{Institute of Systems Science, Durban University of Technology, PO Box 1334,
Durban 4000, RSA}


\begin{abstract}
Scalar-tensor theories have drawn the attention of cosmologist for the past
few years because they can provide mechanism to explain the observable
phenomena. Moreover, the results of scalar-tensor theories can be applied in
higher-order theories of gravity. In this work, we consider a Brans-Dicke
scalar field cosmological model that includes a second scalar field minimally
coupled to gravity. For the second scalar field we consider the two cases that
there exist and there is not any interaction with the Brans-Dicke scalar
field. For the two models of our consideration we prove the existence of
scaling and de Sitter solutions. Furthermore, we perform an Obsiannikov's
classification for the field equations, where we find all the point and
contact transformations which leave the gravitational Action Integral
invariant. Conservation laws are determined and we show how these can be
applied to infer about the integrability of the cosmological models and find
new analytic solutions.

\end{abstract}
\maketitle

\section{Introduction}

\label{sec1}

Scalar fields have been widely used by cosmologists to describe the two major
open problems in the cosmological evolution, namely inflation and dark energy
\cite{cl1,cl2}. Scalar fields are introduced in the Einstein-Hilbert action to
provide new degrees of freedom and modify the dynamics of the field equations
such that tit provides a mechanism which explains the observable phenomena and
the cosmological evolution.

There are various ways to introduce a scalar field in the gravitational Action
Integral. Quintessence is one of the simplest scalar field models where the
scalar field and the gravity are minimally coupled
\cite{ratra,peebles,tsujikawa1}. However, in alternative theories of General
Relativity where the Mach's principle is considered, the existence of the
scalar field is essential for the physical state of gravitational theory and
the scalar field is introduced to be nonminimally coupled with the gravity.
Brans-Dicke theory \cite{Brans} is the most common example of such alternative
gravitational theory. Another alternative is the O'Hanlon theory, which can be
seen as a special case of Brans-Dicke theory. Other example is the Galileon
theory \cite{nik,gal02}. \ These theories belong to a family of a
scalar-tensor theory \cite{faraonibook} also known as the Horndeski theory
\cite{hor}.

Brans-Dicke theory is characterized by a constant $\omega_{BD}$, also known as
Brans-Dicke parameter. For small values of $\omega_{BD}$ the contribution of
the scalar field in the dynamics of the field equations is significant, while
when $\omega_{BD}$ is large, the main contribution in the dynamics is followed
by the tensor part. Hence, if someone expect that as $\omega_{BD}~$reaches
infinity, General Relativity will be recovered, that is not true, which means
that General Relativity and Brans-Dicke gravity are fundamentally different
\cite{omegaBDGR}.

In addition, scalar fields can attribute the degrees of freedom of
higher-order alternative theories of gravity \cite{s01}. For instance, the
fourth-order $f\left(  R\right)  $-gravity is equivalent with the O'Hanlon of
gravity, where the two extra degrees of freedom of $f\left(  R\right)
$-gravity can be described by the O'Hanlon scalar field. The method to perform
such a \textquotedblleft reduction\textquotedblright\ in the order of the
theory by increasing the number of the dependent variables\textquotedblright,
is performed by using a well-known method from analytic Mechanics, the
so-called Lagrange multiplier, some examples on the application of Lagrange
multiplier in cosmological studies can be found in
\cite{noj,s02,s05,s06,s07,s08}.

In \cite{s08} the authors considered a higher-order theory of gravity where
the gravitational action depends on the Ricci scalar and its first and second
derivatives. With the use of Lagrange multipliers, they were able to write the
equivalent scalar tensor theory, which consists of two scalar fields, by using
the latter property the author was able to conclude that this higher-order
theory can be free from Ostrogradsky's instabilities. As far as the two scalar
fields are concerned they can describe a Brans-Dicke field with a second
scalar field minimally coupled to gravity.

In this work, in the context of Brans-Dicke cosmology we consider the
existence of a second scalar field. For the second scalar field we assume that
it is minimally coupled to gravity. As far as the interaction between the two
fields is concerned, we study two cases which lead to two models of study. In
model A, the two fields interact only in the potential part while in the
second model, namely model B, there is a kinetic interaction term for the two
scalar fields in the Action Integral of the theory. These two models are
defined in the Jordan frame. By performing a conformal transformation and
writing the equivalent theories in the Einstein frame we see that the model A
is related with quintom cosmology, while model B with Chiral cosmology, also
known as $\sigma-$model.

The purpose of this work is to determine the exact solutions of these two
models in the Jordan frame, and also to apply the theory of symmetries of
differential equations and determine conservation laws. The latter conserved
quantities can be used to reduce the number of equations, and determine its
feasibility to make a conclusion about the integrability of the field
equations. As we shall see from our analysis, we are able to determine some
integrable models, while we can use the conservation laws to extract
information about the evolution of the scale factor.

In technical terms, the field equations of our cosmological models are three
equations of second-order which form an autonomous Hamiltonian system, where
the Hamiltonian provides the conservation of \textquotedblleft
energy\textquotedblright\ with its value to be constrained. The three
dynamical variables of our models describe the two scalar fields and the scale
factor for the spatially flat Friedmann--Lema\^{\i}tre--Robertson--Walker
(FLRW) which we assume that describes the universe in large scales. We perform
a complete Ovsiannikov-like classification of the scalar field potential of
the models according to the admitted point and contact transformations which
leave invariant the variational principle. The plan of the paper is as follows.

In Section \ref{sec2}, we set the models of our study and we determine the
field equations by using the minisuperspace description. Moreover, we discuss
the equivalency of the two theories of our study in the Einstein frame. In
Section \ref{sec4}, we prove the existence of exact solutions, where the scale
factor is power-law and exponential. These two exact solutions describe
important phases in the evolution of the universe. The determination of
conservation laws for the field equations is presented in Section \ref{sec5}.
Specifically, we present the basic definitions and formulation of variational
symmetries, while we perform a complete classification of the scalar field
potential where the field equations admit conservation laws. Finally, in
Section \ref{conc} we show how the conservation laws can be applied to
construct analytic solutions for the cosmological models of our analysis, and
we draw our conclusions.

\section{Field equations}

\label{sec2}

In this work we consider the Brans-Dicke action \cite{Brans,faraonibook}%
\begin{equation}
S=\int dx^{4}\sqrt{-g}\left[  \frac{1}{2}\phi R-\frac{1}{2}\frac{\omega_{BD}%
}{\phi}g^{\mu\nu}\phi_{;\mu}\phi_{;\nu}+L_{m}\left(  \phi,\psi,\psi_{;\mu
}\right)  \right]  , \label{bd.01}%
\end{equation}
where $\phi\left(  x^{\kappa}\right)  $ is the Brans-Dicke scalar field and
$\omega_{BD}$ is the Brans-Dicke parameter.

Lagrangian function $L_{m}\left(  \phi,\psi,\psi_{;\mu}\right)  $ correspond
to the rest matter source of the model, which is described by a second scalar
field $\psi\left(  x^{\kappa}\right)  $. For the field $\psi\left(  x^{\kappa
}\right)  $ we consider the two scenarios that is: (a) $\psi\left(  x^{\kappa
}\right)  $ is minimally coupled with the Brans-Dicke field, and the Action
Integral (\ref{bd.01}) is written as
\begin{equation}
S_{A}=\int dx^{4}\sqrt{-g}\left[  \frac{1}{2}\phi R-\frac{1}{2}\frac
{\omega_{BD}}{\phi}g^{\mu\nu}\phi_{;\mu}\phi_{;\nu}-\frac{\varepsilon}%
{2}g^{\mu\nu}\psi_{;\mu}\psi_{;\nu}-V\left(  \phi,\psi\right)  \right]  ,
\label{bd.01a}%
\end{equation}
and (b) scalar field $\psi\left(  x^{k}\right)  ,$ is coupled to the
Brans-Dicke field as follows%
\begin{equation}
S_{B}=\int dx^{4}\sqrt{-g}\left[  \frac{1}{2}\phi R-\frac{1}{2}\frac
{\omega_{BD}}{\phi}g^{\mu\nu}\phi_{;\mu}\phi_{;\nu}-\frac{\varepsilon}{2}\phi
g^{\mu\nu}\psi_{;\mu}\psi_{;\nu}-V\left(  \phi,\psi\right)  \right]  .
\label{bd.01b}%
\end{equation}

The two Action Integrals (\ref{bd.01a}), (\ref{bd.01b}) are different,
however, as we shall see in the following section they are related with
well-known cosmological models with two-scalar fields in the Einstein frame.
In particular in the Einstein frame, Action Integral $S_{A}$ becomes $\bar
{S}_{B}$ and correspond to the Chiral cosmology, while $S_{B}$ becomes
$\bar{S}_{B}$ and describe a \textquotedblleft quintom\textquotedblright%
\ cosmological model. The potential form of $V\left(  \phi\left(  x^{\kappa
}\right)  ,\psi\left(  x^{\kappa}\right)  \right)  $ will be defined in the
next section such that exact solutions to derived. Parameter $\varepsilon
^{2}=1$, and plays the role the scalar field $\psi\left(  x^{\kappa}\right)  $
to be quintessence or phantom.

Variations of $S$ (\ref{bd.01}) with respect to the metric tensor gives the
modified Einstein field equations
\begin{equation}
\phi G_{\mu\nu}=\frac{\omega_{BD}}{\phi^{2}}\left(  \phi_{;\mu}\phi_{;\nu
}-\frac{1}{2}g_{\mu\nu}g^{\kappa\lambda}\phi_{;\kappa}\phi_{;\lambda}\right)
-\frac{1}{\phi}\left(  g_{\mu\nu}g^{\kappa\lambda}\phi_{;\kappa\lambda}%
-\phi_{;\mu}\phi_{;\nu}\right)  -g_{\mu\nu}\frac{V\left(  \phi\right)  }{\phi
}+\frac{1}{\phi}T_{\mu\nu},\label{bd.02}%
\end{equation}
where $T_{\mu\nu}$ describes the contribution of the field $\psi\left(
x^{k}\right)  $ in the field equations. In addition, variation with respect to
the fields $\phi\left(  x^{\kappa}\right)  $ and $\psi\left(  x^{\kappa
}\right)  $ provides the second-order differential equations where the two
scalar fields satisfy. 

They are
\begin{subequations}
\begin{equation}
g^{\kappa\lambda}\phi_{;\kappa\lambda}-\frac{1}{2\phi}g^{\mu\nu}\phi_{;\mu
}\phi_{;\nu}+\frac{\phi}{2\omega_{BD}}\left(  R-2V_{,\phi}\right)
=0\label{bd.02a}%
\end{equation}%
\end{subequations}
\begin{equation}
g^{\kappa\lambda}\psi_{;\kappa\lambda}+V_{,\psi}=0\label{bd.02b}%
\end{equation}
for the Action Integral (\ref{bd.01a}), or
\begin{equation}
g^{\kappa\lambda}\phi_{;\kappa\lambda}-\frac{1}{2\phi}g^{\mu\nu}\phi_{;\mu
}\phi_{;\nu}+\frac{\varepsilon\phi}{2\omega_{BD}}g^{\mu\nu}\psi_{;\mu}%
\phi_{;\nu}+\frac{\phi}{2\omega_{BD}}\left(  R-2V_{,\phi}\right)
=0\label{bd.02c}%
\end{equation}%
\begin{equation}
g^{\kappa\lambda}\psi_{;\kappa\lambda}+\frac{1}{\phi}g^{\mu\nu}\phi_{;\mu}%
\psi_{;\nu}+V_{,\psi}=0\label{bd.02d}%
\end{equation}
for the Action Integral (\ref{bd.01b}).

\subsection{Minisuperspace description}

According to the cosmological principle, in large scales the universe is
assumed to be homogeneous and isotropic described by the FLRW spacetime, while
the detailed analysis of the background radiation provides that the spatially
curvature of the spacetime is found to be zero, which means that the line
element which describes the universe in large scales is%

\begin{equation}
ds^{2}=-N^{2}\left(  t\right)  dt^{2}+a^{2}\left(  t\right)  \left(
dx^{2}+dy^{2}+dz^{2}\right)  , \label{bd.05}%
\end{equation}
where $a\left(  t\right)  $ is the scalar factor of the universe and $N\left(
t\right)  $ is the lapse function.

For the line element (\ref{bd.05}) the Ricciscalar is calculated to be
\begin{equation}
R=\frac{6}{N^{2}}\left[  \frac{\ddot{a}}{a}+\left(  \frac{\dot{a}}{a}\right)
^{2}-\frac{\dot{a}\dot{N}}{aN}\right]  . \label{bd.06}%
\end{equation}
where a dot indicates derivative with respect to the independent variable
\textquotedblleft$t$\textquotedblright. \ By replacing (\ref{bd.06}) in the
Action Integrals (\ref{bd.01a}), (\ref{bd.01b}) and integrating by parts we
end up with a point-like Lagrangian whose variation with respect to the
kinematic quantities $N,~a,~\phi$ and $\psi~$produce the gravitational field
equations. At this point we assume that the two scalar fields $\phi\left(
x^{\kappa}\right)  ,~\psi\left(  x^{\kappa}\right)  $ inherits the symmetries
of the spacetime (\ref{bd.05}) from where it follows that $\phi=\phi\left(
t\right)  $ and $\psi=\psi\left(  t\right)  $.

\subsubsection{Point-like Lagrangian for Model A}

From the Action integral (\ref{bd.01a}) and with the use of (\ref{bd.06}) we
find the point-like Lagrangian%

\begin{equation}
L\left(  N,a,\dot{a},\phi,\dot{\phi},\psi,\dot{\psi}\right)  =\frac{1}%
{2N}\left(  -6a\phi\dot{\phi}^{2}-6a^{2}\dot{a}\dot{\phi}-\frac{\omega_{BD}%
}{\phi}a^{3}\dot{\phi}^{2}-\varepsilon a^{3}\dot{\psi}^{2}\right)
+a^{3}NV\left(  \phi,\psi\right)  \label{bd.07}%
\end{equation}
where variation with respect to the kinetic variables $\left\{  a,\phi
,\psi\right\}  ,$ provides the second-order field equations,%
\begin{equation}
0=3\phi H^{2}+2\phi\dot{H}+2H\dot{\phi}-\frac{\omega_{BD}}{2}\dot{\phi}%
^{2}-\frac{\varepsilon}{2}\dot{\psi}^{2}+\ddot{\phi}+V\left(  \phi
,\psi\right)  , \label{bd.08}%
\end{equation}%
\begin{equation}
0=3\dot{H}+6H^{2}+3\omega_{BD}H\frac{\dot{\phi}}{\phi}-\frac{\varepsilon}%
{2}\dot{\psi}^{2}-\frac{\omega_{BD}}{2}\left(  \frac{\dot{\phi}^{2}}{\phi^{2}%
}-2\frac{\ddot{\phi}}{\phi}\right)  +V_{,\phi}\left(  \phi,\psi\right)  ,
\label{bd.09}%
\end{equation}%
\begin{equation}
0=\varepsilon\ddot{\psi}+3\varepsilon H\dot{\psi}+V_{,\psi}\left(  \phi
,\psi\right)  , \label{bd.09a}%
\end{equation}
where we have set $N=1,$ while variation with respect to the lapse function
$N,$ provides the constraint equation $\frac{\partial L}{\partial N}=0$, that
is
\begin{equation}
6\phi H^{2}+6H\dot{\phi}+\frac{\omega_{BD}}{\phi}\dot{\phi}^{2}+\varepsilon
\dot{\psi}^{2}+2V\left(  \phi,\psi\right)  =0. \label{bd.10}%
\end{equation}

For $N=N\left(  a,\phi,\psi\right)  ,~$the field equations (\ref{bd.08}%
)-(\ref{bd.10}) describe the motion of a point particle in the three
dimensional space with line element%
\begin{equation}
ds_{\left(  3A\right)  }^{2}=\frac{1}{N}\left(  -6a\phi d\phi^{2}%
-6a^{2}dad\phi-\frac{\omega_{BD}}{\phi}a^{3}d\phi^{2}+\varepsilon a^{3}%
d\psi^{2}\right)  , \label{bd.11}%
\end{equation}
and effective potential $V_{eff}=a^{3}NV\left(  \phi,\psi\right)  $. The line
element (\ref{bd.11}) is called minisuperspace.

We calculate the Cotton-York tensor
\begin{equation}
C_{\mu\nu\kappa}=R_{\mu\nu;\kappa}-R_{\mu\kappa;\nu}+\frac{1}{4}\left(
R_{;\nu}g_{\mu\kappa}-R_{;\kappa}g_{\mu\nu}\right)  \label{bd.12}%
\end{equation}
for the line element (\ref{bd.11}), from where we find that $C_{\mu\nu\kappa}$
vanishes for arbitrary function $N\left(  a,\phi,\psi\right)  $. Therefore, we
infer that the line element (\ref{bd.11}) describe a three-dimensional
conformally flat spacetime. It is an important property that we shall use in
the following Section in order to determine analytic solutions for the field equations.

\subsubsection{Point-like Lagrangian for Model B}

For the second cosmological model of our consideration with Action Integral
(\ref{bd.01b}) the point-like Lagrangian is found to be%
\begin{equation}
L\left(  N,a,\dot{a},\phi,\dot{\phi},\psi,\dot{\psi}\right)  =\frac{1}%
{2N}\left(  -6a\phi\dot{\phi}^{2}-6a^{2}\dot{a}\dot{\phi}-\frac{\omega_{BD}%
}{\phi}a^{3}\dot{\phi}^{2}-\varepsilon\phi a^{3}\dot{\psi}^{2}\right)
+a^{3}NV\left(  \phi,\psi\right)  \label{bd.13}%
\end{equation}
where now the line element of the three-dimensional minisuperspace which
describes the \textquotedblleft motion\textquotedblright\ of the point-like
particle is%
\begin{equation}
ds_{\left(  3B\right)  }^{2}=\frac{1}{N}\left(  -6a\phi d\phi^{2}%
-6a^{2}dad\phi-\frac{\omega_{BD}}{\phi}a^{3}d\phi^{2}-\varepsilon\phi
a^{3}d\psi^{2}\right)  , \label{bd.14}%
\end{equation}
where again for $N=N\left(  a,\phi,\psi\right)  $ the Cotton-York tensor
$C_{\mu\nu\kappa}$ is calculated to be zero, which means that line element
(\ref{bd.14}) describes a three-dimensional conformally flat space.

From Lagrangian function (\ref{bd.13}) the second-order field equations are
derived to be%
\begin{equation}
0=2\phi\dot{H}+3\phi H^{2}+2H\dot{\phi}-\frac{\omega_{BD}}{2}\frac{\dot{\phi
}^{2}}{\phi}-\frac{\varepsilon}{2}\phi\dot{\psi}^{2}+\ddot{\phi}+V\left(
\phi,\psi\right)  , \label{bd.15}%
\end{equation}%
\begin{equation}
0=3\dot{H}+6H^{2}+3\omega_{BD}H\frac{\dot{\phi}}{\phi}-\frac{\varepsilon}%
{2}\dot{\psi}^{2}-\frac{\omega_{BD}}{2}\left(  \frac{\dot{\phi}^{2}}{\phi^{2}%
}-2\frac{\ddot{\phi}}{\phi}\right)  +V_{,\phi}\left(  \phi,\psi\right)  ,
\label{bd.16}%
\end{equation}%
\begin{equation}
0=\varepsilon\phi\ddot{\psi}+\varepsilon\left(  3H\phi+\dot{\phi}\right)
\dot{\psi}+V_{,\psi}\left(  \phi,\psi\right)  . \label{bd.17}%
\end{equation}
where $H=\frac{1}{N}\frac{\dot{a}}{a}$ is the Hubble function.

Finally, the first-order constraint equation is derived to be%
\begin{equation}
6\phi H^{2}+6H\dot{\phi}+\frac{\omega_{BD}}{\phi}\dot{\phi}^{2}+\varepsilon
\phi\dot{\psi}^{2}+2V\left(  \phi,\psi\right)  =0. \label{bd.18}%
\end{equation}

In what it follows, we investigate the existence of particular solutions for
the two models of our consideration, while we study the integrability of the
dynamical systems by determining conservation laws.

\subsection{Conformal equivalence}

\label{sec3}

We discuss the equivalent conformal theories in the Einstein frame for the two
cosmological models of our consideration. We shall see that in the Einstein
frame the two models (\ref{bd.01a}), (\ref{bd.01b}) becomes the sigma model
\cite{sigm0a,sigm0}, also known as Chiral cosmology \cite{chir3,ndim,anm1} and
the quintom model \cite{quin00,q1,q2,q3,q4}.

Indeed, under the conformal transformation $\bar{g}=\phi g_{ij}$, the Action
integral (\ref{bd.01}) becomes%
\begin{equation}
S=\int dx^{4}\sqrt{-g}\left[  \frac{1}{2}R-\frac{1}{2}g^{\mu\nu}\Phi_{;\mu
}\Phi_{;\nu}+\bar{L}_{m}\left(  \Phi,\psi,\psi_{;\mu}\right)  \right]  ,
\label{bd.018}%
\end{equation}
where $\Phi=\Phi\left(  \phi\left(  x^{k}\right)  \right)  $ is a minimally
coupled scalar field, while $\bar{L}_{m}$ is now the conformal equivalent Lagrangian.

Therefore, the Action Integral (\ref{bd.01a}) which describe model A takes the
form of the Chiral cosmology \cite{anm1}%
\begin{equation}
\bar{S}_{A}=\int dx^{4}\sqrt{-g}\left[  \frac{1}{2}R-\frac{1}{2}g^{\mu\nu}%
\Phi_{;\mu}\Phi_{;\nu}-\frac{\varepsilon}{2}e^{\kappa\Phi}g^{\mu\nu}\psi
_{;\mu}\psi_{;\nu}-\bar{V}\left(  \Phi,\psi\right)  \right]  , \label{bd.019}%
\end{equation}
where constant $\kappa=\kappa\left(  \omega_{BD}\right)  $, while model B
becomes%
\begin{equation}
\bar{S}_{B}=\int dx^{4}\sqrt{-g}\left[  \frac{1}{2}R-\frac{1}{2}g^{\mu\nu}%
\Phi_{;\mu}\Phi_{;\nu}-\frac{\varepsilon}{2}g^{\mu\nu}\psi_{;\mu}\psi_{;\nu
}-\bar{V}\left(  \Phi,\psi\right)  \right]  , \label{bd.020}%
\end{equation}
which is the Action Integral of two scalar field with no interaction in the
kinetic part.

The cosmological models with Action Integrals, the $\bar{S}_{A}$ and $\bar
{S}_{B}$ have been widely studied in the literature and play a significant
role in the description of the evolution of the universe. Quintom models can
provide a dark energy model which cross the cosmological constant boundary,
while model (\ref{bd.019}) is related with the different descriptions of the
inflationary era such as, hyperinflation, $\alpha-$attractors and others
\cite{brown,re1903,atr1}.

Because of the nonlinearity of the field equations, there are very few known
exact solutions, some of them are presented in \cite{ndim,anm1,re1903} and
references therein. While some exact solutions of model (\ref{bd.01a})
presented recently in \cite{ant1,ant2}. The relation between Chiral cosmology
with the modified cosmology is discussed in \cite{verno1}, where some exact
solutions are also presented.

In the latter works a complete analysis on the dynamics of model
(\ref{bd.01a}), for some specific potential forms are presented. From that
analysis it was shown that model (\ref{bd.01a}) can describe various phases of
the cosmological evolution.

\section{Exact solutions}

\label{sec4}

We investigate the existence for solutions of special interest, such as the
scaling solution and the de Sitter universe.

\subsection{Model A}

Before we proceed with our investigation we shall consider special forms of
the scalar field potential $V\left(  \phi,\psi\right)  $. We assume that there
is not any interaction between the two scalar fields, and that $V_{,\phi}=0$.
Hence we assume that $V=V\left(  \psi\right)  $.

\subsubsection{Scaling solution}

We replace $a\left(  t\right)  =a_{0}t^{p},~N\left(  t\right)  =1$, in the
field equations from where we find the second-order ordinary differential
equation%
\begin{equation}
\omega_{BD}\left(  2t\phi\left(  t\ddot{\phi}+6p\dot{\phi}\right)  -t^{2}%
\dot{\phi}^{2}\right)  +6p\left(  5p-1\right)  \phi^{2}=0.\label{bd.19}%
\end{equation}
The solution of the latter equation is%
\begin{equation}
\phi\left(  t\right)  =\phi_{0}t^{1-3p-\Xi\left(  \omega_{BD}\right)  }\left(
t^{\Xi\left(  \omega_{BD}\right)  }+\phi_{1}\right)  ^{2}~,~\omega_{BD}%
\neq0~\ \text{and }p\neq\frac{1}{2}\label{bd.20}%
\end{equation}%
\begin{equation}
\phi\left(  t\right)  =\text{arbitrary,~}p=\frac{1}{2}\text{ and }\omega
_{BD}=0.\label{bd.21}%
\end{equation}
where constant $\Xi\left(  \omega_{BD},p\right)  =\sqrt{\frac{\omega
_{BD}+3p^{2}\left(  3\omega_{BD}-4\right)  -6p\left(  \omega_{BD}-1\right)
}{\omega_{BD}}}$ depends on the Brans-Dicke parameter and the power $p,$ while
$\phi_{0}$ and $\phi_{1}$ are integration constants. We observe that when
$\omega_{BD}=0$, which correspond to O'Hanlon gravity, from (\ref{bd.19}) it
follows that there exists power-law solution iff $p=\frac{1}{2}$ where
$\phi\left(  t\right)  $ is an arbitrary function. We see that for the
radiation solution $a\left(  t\right)  =a_{0}t^{\frac{1}{2}}$, in the case of
O'Hanlon gravity the scalar field $\phi\left(  t\right)  $ plays no role in
the evolution of the universe. Remark that we have assumed $V_{\phi}=0$. This
is a known result, since O'Hanlon gravity, or in the modern consideration,
$f\left(  R\right)  $-gravity provide a radiative phase in the evolution of
the universe when the scalar field potential does not contribute in the field
equations \cite{amen1}. 

Let us consider the radiative solution $a\left(  t\right)  =a_{0}t^{\frac
{1}{2}}$ where $\phi\left(  t\right)  $ is arbitrary. Then from equations
(\ref{bd.10}) and (\ref{bd.08}) it follows that%
\begin{equation}
\dot{\psi}^{2}=\frac{2t^{2}\ddot{\phi}-t\dot{\phi}-2\phi}{2\varepsilon t^{2}},
\label{bd.22}%
\end{equation}%
\begin{equation}
V\left(  \psi\left(  t\right)  \right)  =-\frac{2t^{2}\ddot{\phi}+5t\dot{\phi
}+\phi}{4t^{2}}, \label{bd.23}%
\end{equation}

When $\dot{\psi}\left(  t\right)  =0,$ that is, $\phi\left(  t\right)
=\phi_{1}t^{-\frac{1}{2}}+\phi_{2}t^{2}$, \ the scalar field potential is
derived to be $V\left(  \psi\right)  =-\frac{15}{4}\phi_{2}$.

On the other hand, when $\phi\left(  t\right)  =\phi_{1}t^{-\frac{1}{2}}%
+\phi_{2}t^{2}+\phi_{3}t^{2r}$, we find%
\begin{equation}
\psi\left(  t\right)  =\frac{\sqrt{\left(  r-1\right)  \left(  4r+1\right)  }%
}{r\sqrt{\varepsilon}}~,~V\left(  \psi\left(  t\right)  \right)  =-\frac
{15}{4}\phi_{2}-\frac{1}{4}\left(  1+6r+8r^{2}\right)  \phi_{3}t^{2\left(
r-1\right)  }, \label{bd.24}%
\end{equation}
that is, the scalar field potential $V\left(  \psi\right)  $ is of the form%
\begin{equation}
V\left(  \psi\right)  =V_{0}+V_{1}\left(  \psi-\psi_{0}\right)  ^{2\frac
{\left(  r-1\right)  }{r}}. \label{bd.25}%
\end{equation}
In a similar way we can construct another kind of potentials.

Consider solution (\ref{bd.20}) and for simplicity on our calculations, we
assume $\phi_{1}=0$ such that $\phi\left(  t\right)  =\phi_{0}t^{1-3p+2\Xi
\left(  \omega_{BD},p\right)  }$. Thus for the scalar field $\psi\left(
t\right)  $ we find,
\begin{equation}
\dot{\psi}^{2}=\frac{\left(  p-1-\Xi\right)  \left(  1+3p-\Xi\right)  \Xi
}{\varepsilon\left(  3p-1+\Xi\right)  }\phi_{0}t^{-1-3p+\Xi}, \label{bd.26}%
\end{equation}
while the scalar field potential is given as a function of $``t$%
\textquotedblright,\ as follows%
\begin{equation}
V\left(  \psi\left(  t\right)  \right)  =\frac{\phi_{0}}{2}\Xi\left(
p-1-\Xi\right)  t^{-1+3p+\Xi}. \label{bd.27}%
\end{equation}

Finally, with the use of (\ref{bd.26}) the potential as a function of the
scalar field is found to be%
\begin{equation}
V\left(  \psi\right)  =V_{0}+V_{1}\left(  \psi-\psi_{0}\right)  ^{2-\frac
{4}{1-3p+\Xi}}. \label{bd.28}%
\end{equation}
Note that condition $1-3p+\Xi=0$ indicates that $\omega_{BD}\rightarrow\infty
$, hence solution (\ref{bd.28}) is valid for finite values of the Brans-Dicke
parameter $\omega_{BD}$.

\subsubsection{de Sitter Solution}

Without loss of generality, we set $N\left(  t\right)  =\frac{1}{H_{0}}a$, and
$a=t$. The latter selection provides a de Sitter universe with expansion rate
$H_{0}$. Thus, for the Brans-Dicke scalar field we find%
\begin{equation}
\phi=\phi_{0}t^{-3-\sqrt{9-\frac{12}{\omega_{BD}}}}\left(  t^{\sqrt
{9-\frac{12}{\omega_{BD}}}}+\phi_{1}\right)  ^{2},\label{bd.29}%
\end{equation}
where again for simplicity on our calculations we assume $\phi_{1}=0$. 

For the minimally coupled scalar field it follows%
\begin{equation}
\dot{\psi}^{2}=\phi_{0}\frac{\left(  6\omega_{BD}^{2}\left(  \sqrt{9-\frac
{12}{\omega_{BD}}}-3\right)  -7\omega_{BD}\sqrt{9-\frac{12}{\omega_{BD}}%
}-12\right)  }{\varepsilon\omega_{BD}}t^{-5+\sqrt{9-\frac{12}{\omega_{BD}}}%
}\label{bd.30}%
\end{equation}%
\begin{equation}
V\left(  \psi\left(  t\right)  \right)  =\frac{\phi_{0}}{2\omega_{BD}}\left(
12+\left(  \sqrt{9-\frac{12}{\omega_{BD}}}-9\right)  \omega_{BD}\right)
t^{-3+\sqrt{9-\frac{12}{\omega_{BD}}}}\label{bd.31}%
\end{equation}
that is, the scalar field potential is determined to be%
\begin{equation}
V\left(  \psi\right)  =V_{0}+V_{1}\psi^{2}.\label{bd.32}%
\end{equation}

Before we proceed with the next model of our consideration, we conclude that
exact solutions exist for the power law scalar field model $V\left(
\psi\right)  =V_{0}+V_{1}\psi^{P}~$where for $P=2$, the exact solution is the
nonsingular de Sitter solution, while for $P\neq2$ the spacetime has an
initial singularity.

\subsection{Model B}

For the unknown potential in the Action Integral (\ref{bd.01b}), we assume
that $V\left(  \phi,\psi\right)  =\phi V\left(  \psi\right)  $.

\subsubsection{Scaling solution}

For the latter selection of the scalar field potential and for scale factor of
the form $a\left(  t\right)  =a_{0}t^{\frac{p-1}{3}}$ with $N\left(  t\right)
=1$, from the system (\ref{bd.15})-(\ref{bd.18}), we find the solution for the
Brans-Dicke scalar field%
\begin{equation}
\phi\left(  t\right)  =\phi_{1}t^{\Xi_{+}}+\phi_{2}t^{\Xi_{-}} \label{bd.33}%
\end{equation}
where%
\begin{equation}
\Xi_{\pm}\left(  \omega_{BD,}p\right)  =\frac{1}{6}\left(  \left(  \frac
{1}{\omega_{BD}-1}-3\right)  p+\frac{1}{\omega_{BD}-1}\right)  \pm
\frac{p\left(  3\omega_{BD}-4\right)  -1}{\omega_{B}-1}~,~\omega_{BD}\neq1.
\label{bd.34}%
\end{equation}

By considering $\phi_{1}=0$ or $\phi_{2}=0$, for the scalar field $\psi\left(
t\right)  $ the closed-form solution is as follows,%
\begin{equation}
\psi\left(  t\right)  =\psi_{0}\left(  p,\phi_{\left(  1,2\right)  }%
,\omega_{BD}\right)  \ln t~,~V\left(  \psi\left(  t\right)  \right)  \simeq
t^{-2} \label{bd.35}%
\end{equation}
from where we can infer that the scalar field potential is exponential
function of $\psi$, that is%
\begin{equation}
V\left(  \psi\right)  =V_{0}\left(  p,\phi_{\left(  1,2\right)  },\omega
_{BD}\right)  e^{-\lambda\psi},~\lambda=\lambda\left(  p,\omega_{BD}\right)  .
\label{bd.36}%
\end{equation}

For $\omega_{BD}=1$, the exact solution of the field equations is
\begin{equation}
\phi\left(  t\right)  =\phi_{0}t^{-p}~,~\psi\left(  t\right)  =\sqrt
{\frac{\left(  p+1\right)  ^{2}-3}{3\varepsilon}}\ln t~,~V\left(  \psi\right)
=0.
\end{equation}
Remark that $\omega_{BD}=1$ corresponds to the Dilaton field which is
invariant under the duality transformation.

\subsubsection{de Sitter Solution}

In order to investigate the existence of a de Sitter universe, we follow the
same step as we did for model A. We replace $N\left(  t\right)  =\frac
{1}{H_{0}a},~a\left(  t\right)  =t$ in the field equations (\ref{bd.15}%
)-(\ref{bd.18}) where we find,%
\begin{equation}
\phi\left(  t\right)  =\phi_{0}t^{\Phi\left(  \omega_{BD}\right)  ~}%
,~\psi\left(  t\right)  =\psi_{0}\left(  \phi_{0},\omega_{BD}\right)  \ln
t~,~V\left(  \psi\right)  =V_{0}\left(  \omega_{B},H_{0}\right)  ,~\omega
_{BD}\neq1
\end{equation}
or%
\begin{equation}
\phi\left(  t\right)  =\phi_{0}t^{-3},~\psi\left(  t\right)  =\sqrt{\frac
{3}{\varepsilon}}\ln t~,~V\left(  \psi\right)  =0.
\end{equation}

Therefore, we conclude that for the scalar field potential $V\left(  \phi
,\psi\right)  =V_{0}\phi e^{-\lambda\psi},$ there exist a scaling solution for
$\lambda\neq0$, while de Sitter solution exists only when $\lambda=0$. For the
special case of $\omega_{BD}=1$, those exact solutions exists if and only if
$V_{0}=0$, that is, the scalar field potential is zero, $V\left(  \phi
,\psi\right)  =0$.

Until now we investigate the existence of some exact solutions for the models
of our consideration. The solutions we studied are special solutions and they
do not describe the general cosmological evolution of our models.

The models that we consider consist of three equations of second-order which
means their degree of freedom is six and we need to determine at least three
conservation laws in order to conclude about the integrability of the models,
consequently to prove the existence of actual solutions.

This task is the subject of Section \ref{sec5}, where technics from the study
of the integrability of Hamiltonian systems are applied to prove the
integrability of these nonlinear cosmological models.

\section{Symmetries and integrability}

\label{sec5}

There are various ways to describe the meaning of integrability for a
dynamical system. Indeed, when there exist a set of explicit functions which
describe the variation of the dependent variables with the independent
variable, we shall say that there exists a solution for the dynamical system,
as we saw in the previous section. However, in order to infer if the dynamical
system is integrable, and admits an actual solution for arbitrary initial
conditions, the set of explicit functions which describe the variation of the
dependent variables with the independent variable for the dynamical system
should have a number of arbitrary constants equal with the degree of freedom
of the dynamical systems. That constants, are also called integration
constants and specifically they describe conservation laws for the dynamical system.

There is not unique method to construct conservation laws of dynamical
systems. Some of the different approaches which have been applied in
cosmological studies are, (a) point symmetries (b) dynamical symmetries
\cite{dyn3,dyn4}, (c) Cartan symmetries \cite{car1}, (d) nonlocal symmetries
\cite{non1,non2,non3,non4} (e) singularity analysis \cite{sin1}, (f) Hojman
approach \cite{hoj,hoj2}, (g) Darboux polynomial \cite{dar1,dar2}, (h)
automorphism \cite{tc1,tc2,tc3}, (i) Galois group \cite{gal1}.

In gravitational theories with a minisuperspace description, the application
of symmetries has been mainly performed by using Noether's theorems
\cite{noe}. Moreover, Noether's theorem applied as geometric criterion to
classify various cosmological models according to their group properties,
similarly with Ovsiannikov's classification \cite{obsi}. From the latter
classification scheme new integrable models and analytic solutions have been
determined in the literature, for instance see
\cite{ns1,ns2,ns3,ns4,ns5,ns6,ns7,ns8,ns9} and references therein.
Furthermore, symmetries have been used and in other aspects of the theories,
like in quantum cosmology and others \cite{aps1,aps2,aps3}.

In this work, we apply Noether's theorem to determine conservation laws for
the two cosmological models of our analysis, and study the integrability of
the field equations. We consider the existence of point transformations and
dynamical transformations which leaves invariant the variation of the Action
integrals (\ref{bd.01a}) and (\ref{bd.01b}). For the convenience of the reader
we briefly discuss the basic properties and definitions on the application of
symmetries on the variational principle.

\subsection{Preliminaries}

We briefly discuss the main mathematical tools which are applied in this work
for the study of the integrability of the cosmological models (\ref{bd.01a})
and (\ref{bd.01b}).

\subsubsection{Noether's theorems}

Let \thinspace$S\left(  q^{\mu},\dot{q}^{\mu}\right)  $ is the Action Integral%
\begin{equation}
S\left(  q^{\mu},\dot{q}^{\mu}\right)  =\int_{t_{1}}^{t_{2}}L\left(  t,q^{\mu
},\dot{q}^{\mu}\right)  dt. \label{FuncAction}%
\end{equation}
whose variational provide the equations of motion for a given dynamical system.

In the jet space, $J^{1}\left(  M\right)  $ we define the vector field
$X_{N}=X^{\left[  1\right]  }$ which is the generator of the infinitesimal
transformation
\begin{align}
t^{\prime}  &  =t+\varepsilon\xi\left(  t,q^{\nu},...\right)  ,\label{tr1}\\
q^{\mu^{\prime}}  &  =q^{i}+\varepsilon\eta^{\mu}\left(  t,q^{\nu},...\right)
,\\
\dot{q}^{\mu^{\prime}}  &  =\dot{q}^{\mu}+\varepsilon\left(  \dot{\eta}^{\mu
}-\dot{\xi}\dot{q}^{\mu}+\phi^{i}\right)  . \label{TransfHol}%
\end{align}
and~$X^{\left[  1\right]  }=\xi\partial_{t}+\eta^{\mu}\partial_{\mu}+\dot
{\eta}^{\left[  1\right]  }\partial_{\dot{q}^{\mu}}$.

The vector field $X_{N}$ is called Noether symmetry of the dynamical described
by the Action Integral (\ref{FuncAction}) iff%
\begin{equation}
S\left(  q^{\mu},\dot{q}^{\mu}\right)  =S\left(  q^{\mu},\dot{q}^{\mu}\right)
+\epsilon\int_{t_{1}}^{t_{2}}\frac{df(t,q^{\mu},\dot{q}^{\mu})}{dt}dt
\label{con1}%
\end{equation}
where $f(t,q^{\mu},\dot{q}^{\mu})$ is a smooth function and the infinitesimal
transformation causes a zero end point variation (i.e. the end points of the
integral remain fixed).

From (\ref{con1}) it follows that the equation \cite{sarlet},%
\begin{equation}
X^{\left[  1\right]  }\left(  L\right)  +L\left(  t,q^{i},\dot{q}^{i}\right)
\dot{\xi}=\dot{f} \label{GenHolonNoether's1Cond}%
\end{equation}
which is called Noether symmetry condition, or Noether's first theorem.

With the use of (\ref{GenHolonNoether's1Cond}) and of the equations of motion
$\frac{d}{dt}\left(  \frac{\partial L}{\partial\dot{q}^{\mu}}\right)
-\frac{\partial L}{\partial q^{\mu}}=0$, we find
\begin{equation}
0=\frac{d}{dt}\left(  f-L\xi-\frac{\partial L}{\partial\dot{q}^{i}}\left(
\eta^{i}-\xi\dot{q}^{i}\right)  \right)  . \label{GenHolonNoether'2 Cond}%
\end{equation}
which leads to the first integral of the dynamical system
\begin{equation}
I=f-L\xi-\frac{\partial L}{\partial\dot{q}^{i}}\left(  \eta^{i}-\xi\dot{q}%
^{i}\right)  , \label{GenHolonNoether's1Integr}%
\end{equation}
also known as Noether's second theorem.

Noether's theorems are valid not only for point symmetries but also for
generalized symmetries, and contact symmetries. For point symmetries, where
$\xi=\xi\left(  t,q^{\nu}\right)  $ and $\eta^{\mu}=\eta^{\mu}\left(
t,q^{\nu}\right)  ,$ function $\xi$ is essential in-order to derive the full
group of Noether symmetries for a given dynamical system. However, for contact
symmetries (or also known as dynamical symmetries) or for generalized
symmetries, where $\eta^{\mu}$ is a non-constant function of $\dot{q}^{\nu}%
$,$~$function $\xi$ can be always omitted \cite{Kalotas}; thus Noether
condition (\ref{GenHolonNoether's1Cond}) is simplified as%
\begin{equation}
X^{\left[  1\right]  }\left(  L\right)  -\dot{f}=0, \label{bd.40}%
\end{equation}
while Noether's second theorem is written as follows%
\begin{equation}
I=f-\frac{\partial L}{\partial\dot{q}^{i}}\eta^{i}. \label{bd.41}%
\end{equation}

\subsection{Conservation laws for Model A}

For the cosmological model with point-like Lagrangian (\ref{bd.07}), we assume
that $V\left(  \phi,\psi\right)  =V\left(  \psi\right)  $ is non-constant
function and the lapse function to be $N=1$.

We consider the generator $X_{P}$ of the one-parameter point transformation,%
\begin{align}
t^{\prime}  &  =t+\varepsilon\xi\left(  t,a,\phi,\psi\right)  ~,~a^{\prime
}=a+\varepsilon\eta^{a}\left(  t,a,\phi,\psi\right) \label{bd.42}\\
\phi^{\prime}  &  =\phi+\varepsilon\eta^{\phi}\left(  t,a,\phi,\psi\right)
~,~\psi^{\prime}=\psi+\varepsilon\eta^{\psi}\left(  t,a,\phi,\psi\right)
\label{bd.43}%
\end{align}
and we apply condition (\ref{GenHolonNoether's1Cond}) with Lagrangian
(\ref{bd.07}).

From the symmetry condition (\ref{GenHolonNoether's1Cond}) it follows a system
of linear partial differential equations with dependent variables the
$\xi,~\eta^{a},~\eta^{\phi},~\eta^{\psi}$ and $V\left(  \psi\right)  \,$. The
solution of the later system specify the specific forms of $V\left(
\psi\right)  $ where symmetry vectors exist.

Recall that the field equations form an autonomous dynamical system which
admits the Noether symmetry vector $\partial_{t}$, and corresponding
conservation law the Hamiltonian, which is actually\ the constraint equation
(\ref{bd.18}).

For $\omega_{BD}\neq0$, the unique scalar field potential which admits the
additional point symmetry $X_{P}^{1}=2t\partial_{t}+Y_{P}^{1},~Y_{P}^{1}%
=\frac{2}{3}a\partial_{a}-2\phi\partial_{\phi}-\psi\partial_{\psi},$ for the
quadratic potential $V\left(  \psi\right)  =V_{0}\left(  \psi-\psi_{0}\right)
^{2},$ where without loss of generality we set $\psi_{0}=0$. Furthermore, from
formula (\ref{GenHolonNoether's1Integr}) we can write the corresponding
conservation law,%
\begin{equation}
I_{1}=10a^{2}\dot{a}\dot{\phi}+2a^{3}\left(  1+\omega_{BD}\right)  \dot{\phi
}+\varepsilon a^{3}\psi\dot{\psi}. \label{bd.44}%
\end{equation}

Nevertheless, the conservation laws that we have derived for the model, are
less from the number of the degrees of freedom of the model. Hence, the
existence of a third conservation law should be studied in order to conclude
for the integrability of this model.

We continue by extending the content of symmetries to the case of contact
symmetries/dynamical symmetries where now $X_{G}$ is the generator of the
infinitesimal transformation,%
\begin{align}
t^{\prime}  &  =t~,~a^{\prime}=a+\varepsilon\left(  \eta_{a}^{a}\left(
a,\phi,\psi\right)  \dot{a}+\eta_{\phi}^{a}\left(  a,\phi,\psi\right)
\dot{\phi}+\eta_{\psi}^{a}\left(  a,\phi,\psi\right)  \dot{\psi}\right)
~,\label{bd.45}\\
\phi^{\prime}  &  =\phi+\varepsilon\left(  \eta_{a}^{\phi}\left(  a,\phi
,\psi\right)  \dot{a}+\eta_{\phi}^{\phi}\left(  a,\phi,\psi\right)  \dot{\phi
}+\eta_{\psi}^{\phi}\left(  a,\phi,\psi\right)  \dot{\psi}\right)
~,\label{bd.45a}\\
,~\psi^{\prime}  &  =\psi+\varepsilon\left(  \eta_{a}^{\psi}\left(
a,\phi,\psi\right)  \dot{a}+\eta_{\phi}^{\psi}\left(  a,\phi,\psi\right)
\dot{\phi}+\eta_{\psi}^{\psi}\left(  \psi a,\phi,\psi\right)  \dot{\psi
}\right)  . \label{bd.46}%
\end{align}
From the theory of differential equations we know that $X_{G}$ is a Noether
symmetry, if the two dimensional tensor $\eta_{~j}^{i}\left(  a,\phi
,\psi\right)  $ is a Killing tensor for the three-dimensional minisuperspace
(\ref{bd.14}), and an additional condition for the potential is satisfied.

For $\omega_{BD}\neq0$ we find that for arbitrary potential $V\left(
\psi\right)  $, the unique Killing tensor which provide a Noether point
symmetry is the metric tensor (\ref{bd.14}) with corresponding conservation
law the constraint equation (\ref{bd.18}).

For $V\left(  \psi\right)  =V_{0}\psi^{2}$, there exists a second Killing
tensor which provide a contact symmetry, which is the vector product of
$Y_{p}^{1}\wedge Y_{P}^{1}$, where the corresponding conservation law is
$I_{2}^{G}=\left(  I_{1}\right)  ^{2}.$ Consequently, we can not infer for the
integrability of the cosmological model. \ However, for Brans-Dicke parameter
zero the results are different.

In particular for $\omega_{BD}=0,$ for arbitrary potential $V\left(
\psi\right)  $, the field equations are invariant under the action of the
point transformation $X_{P}^{2}=\frac{1}{a}\partial_{\phi}$, with
corresponding conservation law $I_{2}=a\dot{a}.$

From the latter conservation law it follows the radiation solution, $a\left(
t\right)  =a_{0}t^{\frac{1}{2}}$, which was found in the previous section. In
addition for $V\left(  \psi\right)  =V_{0}\psi^{2},~$the field equations are
invariant under the action of the Noether point symmetries $X_{P}^{1}$ with
conservation law (\ref{bd.44}).

We search for contact symmetries and we find that for arbitrary $V\left(
\psi\right)  $, the Killing tensors which provide conservation laws are the
metric tensor and the $X_{p}^{2}\wedge X_{P}^{2}$ with \ the corresponding
conservation laws being the constraint equation (\ref{bd.14}) and $I_{3}%
^{G}=\left(  I_{2}\right)  ^{2}.$ For the power-law potential $V\left(
\psi\right)  =V_{0}\psi^{2}$, there exist two additional Killing tensors which
satisfy the symmetry condition (\ref{GenHolonNoether's1Cond}), they are
$Y_{p}^{1}\wedge Y_{P}^{1}$ and $Y_{p}^{1}\wedge X_{P}^{2}$, where the
corresponding conservation laws are~$I_{2}^{G}$ and $I_{4}^{G}=I_{1}I_{2}.$

Furthermore, for $\omega_{BD}=0$ and $V\left(  \psi\right)  =V_{0}\psi
^{K},~K\neq0,2$ the dynamical system is invariant under the contact
transformation (\ref{bd.45})-(\ref{bd.46}) generated by the KTs with
components%
\begin{equation}
\eta_{aa}=-2\left(  K-10\right)  \phi a^{3}~,~\eta_{a\phi}=\left(
K-10\right)  a^{4}~,~~\eta_{a\psi}=4\varepsilon a^{4}\psi~,~\eta_{\psi\psi
}=\varepsilon\frac{K-2}{K-10}a^{5}, \label{bd.47}%
\end{equation}
and boundary term $f=a^{5}\psi^{K}$.

Finally, the corresponding conservation law is calculated from formula
(\ref{GenHolonNoether's1Integr}), that is,
\begin{equation}
I_{4}^{G}=\left(  K-10\right)  \left(  a^{4}\dot{a}\dot{\phi}-\phi a^{3}%
\dot{a}^{2}\right)  +4\varepsilon\psi a^{4}\dot{a}\dot{\psi}+\frac{K-2}%
{2}\varepsilon a^{5}\dot{\psi}^{2}+\left(  K-2\right)  a^{5}\psi^{K}.
\label{bd.48}%
\end{equation}

\subsubsection{Potential $V_{,\phi}\neq0$}

An open question of that specific classification scheme is what happens with
the conservation laws when the scalar field potential is also a function of
the Brans-Dicke field, that is $V\left(  \phi,\psi\right)  _{,\phi}\neq0$. As
far as the point symmetries is concerned, we can give the following answer.

For $\omega_{BD}\neq0$, and $V_{1}\left(  \phi,\psi\right)  =F_{1}\left(
\frac{\psi^{2}}{\phi}\right)  \phi$, the field equations admit as (Noether)
point symmetry the vector field $X_{P}^{1}$. As we can see when \thinspace
$F_{1}\left(  \frac{\psi^{2}}{\phi}\right)  =\frac{\psi^{2}}{\phi}$, then we
recover the previous results. For the latter potential the only second-rank
Killing tensor which define a contact symmetry is the $Y_{p}^{1}\wedge
Y_{P}^{1}$ with conservation law $I_{2}^{G}.$

In the case of O'Hanlon gravity, i.e. $\omega_{BD}=0$, the results are
different. Indeed, as far as the classification of point symmetries is
concerned, we have the two potentials $V_{1}\left(  \phi,\psi\right)
=F_{1}\left(  \frac{\psi^{2}}{\phi}\right)  \phi$ and $V_{2}\left(  \phi
,\psi\right)  =F_{1}\left(  \varepsilon\frac{\psi^{2}}{6}-\phi\right)  $ and
$V_{3}\left(  \phi,\psi\right)  =V_{0}\phi-V_{0}\frac{\varepsilon}{6}\psi^{2}%
$. \ The point symmetries of potential $V_{1}\left(  \phi,\psi\right)  $ are
exactly that for $\omega_{BD}\neq0$, however for the two new potentials,
namely $V_{2}\left(  \phi,\psi\right)  $ and $V_{3}\left(  \phi,\psi\right)  $
new (Noether) point symmetries exist for the field equations.

For the potential $V_{2}\left(  \phi,\psi\right)  ,$ the field equations are
invariant under the one-parameter point transformation (\ref{bd.42}),
(\ref{bd.43}) with generators the vector field $X_{P}^{3}=\frac{\varepsilon
}{3}\frac{\psi}{a}\partial_{\phi}+\frac{1}{a}\partial_{\psi}~$while the extra
conservation law is derived to be%
\begin{equation}
I_{3}^{P}=\psi a\dot{a}+a^{2}\dot{\psi}. \label{bd.49}%
\end{equation}

For the potential $V_{3}\left(  \phi,\psi\right)  $ is actually a particular
case of $V_{1}\left(  \phi,\psi\right)  ~$and $V_{2}$, then the point
symmetries are $X_{P}^{1}$ and $X_{P}^{2}$ and $X_{P}^{3}$ with conservation
laws the $I_{1}$~,$~I_{2}$ and $I_{3}.$

We proceed with the second cosmological model of our consideration.

\subsection{Conservation laws for Model B}

For the point-like Lagrangian (\ref{bd.13}), the complete classification of
the transformations of the form (\ref{bd.42})-(\ref{bd.43}), and
(\ref{bd.45})-(\ref{bd.46}), which satisfy the symmetry condition
(\ref{GenHolonNoether's1Cond}) is presented in the following lines.

\subsubsection{Conservation laws from point symmetries}

The field equations (\ref{bd.15})-(\ref{bd.18}) are invariant under the
following point symmetries.

For arbitrary potential $V\left(  \phi,\psi\right)  $, the dynamical system is
autonomous and admits the trivial Noether symmetry the vector field
$\partial_{t}$ with conservation law the constraint equation (\ref{bd.18}).

For $V_{1}\left(  \phi,\psi\right)  =\Phi_{1}\left(  \phi\right)  $, the
dynamical system admits the extra point symmetry $Z_{1}=\partial_{\psi},$ with
conservation law $\Sigma_{1}=a^{3}\phi\dot{\psi}$. In the special case where
$\Phi_{1}\left(  \phi\right)  =\Phi_{0}\phi,$ is a linear function the
dynamical system admits the additional symmetry vectors,%
\begin{align}
Z_{2}  &  =-\frac{a}{3}\partial_{a}+\phi\partial_{\phi},\\
Z_{3}  &  =a\psi-3\psi\phi+\frac{3}{\varepsilon}\left(  \left(  \omega
_{BD}-1\right)  \ln\phi+\ln\left(  a\right)  \right)  \partial_{\psi},
\end{align}
with conservation laws%
\begin{align}
\Sigma_{2}  &  =a^{2}\phi\dot{a}+\left(  \omega_{BD}-1\right)  a^{3}\dot{\phi
},\\
\Sigma_{3}  &  =a^{2}\phi\psi\dot{a}+a^{3}\psi\left(  \omega_{BD}-1\right)
\psi a^{3}\dot{\phi}-\phi a^{3}\left(  \left(  \omega_{BD}-1\right)  \ln
\phi+\ln\left(  a\right)  \right)  \dot{\psi},
\end{align}

Moreover, when the scalar field potential is $V_{2}\left(  \phi,\psi\right)
=\Phi_{2}\left(  \phi^{K}e^{\psi}\right)  \phi$, the field equations admit the
Noether symmetry $Z_{4}=Z_{2}+KZ_{1},$ with conservation law $\Sigma
_{4}=\Sigma_{2}+K\Sigma_{1}$. In addition, when~$V_{3}\left(  \phi
,\psi\right)  =V\left(  \phi^{\frac{K}{3\lambda-2}}e^{\psi}\right)
\phi^{\frac{3\lambda+2}{3\lambda-2}}$ the Noether point symmetries are $Z_{4}$
and $Z_{5}=2t\partial_{t}+\phi\partial_{\phi}$ with conservation laws
$\Sigma_{4}$ and $\Sigma_{5}=a^{2}\left(  2\phi\dot{a}+a\dot{\phi}\right)  $.

Finally, for $V_{4}\left(  \phi,\psi\right)  =\Phi_{4}\left(  \phi\right)
e^{-\lambda\psi}$, the field equations admit the additional Noether point
symmetry $Z_{6}=2t\partial_{t}+\frac{2}{3}a\partial_{a}+\frac{1}{\lambda
}\partial_{\psi},$ with conservation law $\Sigma_{6}=a^{2}\left(  3\phi\dot
{a}+\omega_{BD}a\dot{\phi}\right)  $.

\subsubsection{Conservation laws from contact symmetries}

We continue by presenting the classification of the scalar field potentials
and of the corresponding conservation laws for contact symmetries. We omit the
presentation of the conservation laws which reduced to those of the point symmetries.

For $V_{5}\left(  \phi,\psi\right)  =V_{0}\phi+\left(  V_{1}e^{2\sqrt
{3\varepsilon\left(  3\omega_{BD}-4\right)  }\psi}+V_{2}e^{-2\sqrt
{3\varepsilon\left(  3\omega_{BD}-4\right)  }\psi}\right)  \phi^{6\omega
_{BD}-7}$, the field equations admit the conservation law,%
\begin{equation}
\Sigma_{7}=\phi^{2}a^{5}\dot{a}\dot{\phi}+\left(  \omega_{BD}-1\right)
a^{6}\phi\dot{\phi}\dot{\psi}-\frac{\sqrt{3\varepsilon}}{3}a^{6}\phi^{6\left(
\omega_{BD}-1\right)  }\left(  V_{1}e^{2\sqrt{3\varepsilon\left(  3\omega
_{BD}-4\right)  }\psi}-V_{2}e^{-2\sqrt{3\varepsilon\left(  3\omega
_{BD}-4\right)  }\psi}\right)  .
\end{equation}

When $V_{6}\left(  \phi,\psi\right)  =V_{0}\phi+V_{1}\phi^{6\omega_{BD}-7}$,
then there exists the extra quadratic conservation law,%
\begin{equation}
\Sigma_{8}=a^{5}\phi\dot{a}\dot{\phi}+\frac{\left(  \omega_{BD}-1\right)  }%
{2}a^{6}\dot{\phi}^{2}+\frac{a^{4}\phi^{2}}{2\left(  \omega_{BD}-1\right)
}\dot{a}^{2}+V_{1}\frac{a^{6}\left(  3\omega_{BD}-4\right)  }{3\left(
\omega_{BD}-1\right)  }\phi^{6\left(  \omega_{BD}-1\right)  }.
\end{equation}
Recall that $V_{6}\left(  \phi,\psi\right)  $ is a special case of
$V_{1}\left(  \phi,\psi\right)  $, that is, the point symmetry $Z_{1}$ also exists.

For the scalar field potential $V_{7}\left(  \phi,\psi\right)  =V_{0}\phi
+\Phi_{7}\left(  \psi\right)  \phi^{6\omega_{BD}-7}$, the field equations
admit the quadratic conservation law,%
\begin{equation}
\Sigma_{9}=a^{5}\phi\dot{a}\dot{\phi}+\frac{\left(  \omega_{BD}-1\right)  }%
{2}a^{6}\dot{\phi}^{2}+\frac{a^{4}\phi^{2}}{2\left(  \omega_{BD}-1\right)
}\dot{a}^{2}+\frac{3\omega_{BD}-4}{6\left(  \omega_{BD}-1\right)  }a^{6}%
\phi^{2}\dot{\psi}^{2}+\frac{\left(  3\omega_{BD}-4\right)  }{3\left(
\omega_{BD}-1\right)  }a^{6}\Phi_{7}\left(  \psi\right)  \phi^{6\left(
\omega_{BD}-1\right)  }.
\end{equation}

Finally for the scalar field potential $V_{8}\left(  \phi,\psi\right)  =\phi
V_{8}\left(  \psi\right)  +V_{1}\phi^{6\omega_{BD}-7}$, the dynamical system
(\ref{bd.15})-(\ref{bd.18}) admits the extra quadratic conservation law
$\Sigma_{8}.$

\section{Conclusions}

\label{conc}

For the two cosmological models of our consideration, we calculated all the
possible conservation laws generated by the application of Noether's first and
second theorems for point and contact transformations. By using the
conservation laws we are able to determine exact solutions or infer about the
integrability of the dynamical system.

In addition, we show that the exact solutions which derived in Section
\ref{sec4}, are directly related with the admitted conservation laws. As we
discussed above for the Model A, and for $\omega_{BD}=0,~V\left(  \phi
,\psi\right)  =V_{0}\psi^{2},~$the admitted conservation law $I_{2}=a\dot{a},$
provides the radiation solution $a\left(  t\right)  =a_{0}t^{\frac{1}{2}}$.

Consider now Model B with scalar field potential $V_{1}\left(  \phi\right)
=V_{0}\phi^{2}.$ By using the conservation law $\Sigma_{1}$ we find $\psi
=\psi_{0}+\int\Sigma_{1}a^{-3}\phi^{-1}dt$, while from $\Sigma_{8}$ it follows
$a\left(  t\right)  =a_{0}\left(  \phi^{^{\prime}}\right)  ^{-1/3},$ where the
field equations are reduced to the third-order ordinary differential equation,%
\begin{equation}
\phi\left(  \dot{\phi}\phi^{\left(  3\right)  }-2\dot{\phi}^{2}\right)
+\dot{\phi}^{2}\ddot{\phi}-3V_{0}\phi^{2}\dot{\phi}^{2}=0.\label{bd.99}%
\end{equation}

For the latter ordinary differential equation we perform the change of
variable $\phi=\phi_{0}+\frac{1}{\varpi\left(  t\right)  }$. In order to solve
the equation in terms of $\varpi\left(  t\right)  $ we apply the singularity
analysis \cite{sin1} and we find the analytic solution which is expressed in
terms of the Laurent expansion as follows%
\begin{equation}
\varpi\left(  t\right)  =\varpi_{0}\left(  t-t_{0}\right)  ^{-1}+\varpi_{1}+%
{\displaystyle\sum\limits_{i=2}}
\varpi_{i}\left(  t-t_{0}\right)  ^{-1+i},\label{bd.100}%
\end{equation}
where $\varpi_{0},~\varpi_{1}$ and $t_{0}$ are the three integration constants
of the solution and coefficients $\varpi_{i}=\varpi_{i}\left(  \varpi
_{0},\varpi_{1},V_{0}\right)  $.

Therefore, the scale factor is expressed as $a\left(  t\right)  \simeq\left(
\frac{\varpi^{\prime}}{\varpi^{2}}\right)  ^{-1/3},$ where for small values of
$t-t_{0},$ the scale factor is approximated to be $a\left(  t\right)
\simeq\left(  1+\frac{\varpi_{1}}{\varpi_{0}}\left(  t-t_{0}\right)  \right)
^{\frac{2}{3}}$, where for small values of the ration $\frac{\varpi_{1}%
}{\varpi_{0}}$ the latter solution describes the matter dominated era
$a\left(  t\right)  \simeq t^{\frac{2}{3}}$.

On the other hand, equation (\ref{bd.99}) can be integrated by quadratures.
Indeed under the similarity transformation $u=\dot{\phi}$,~$v=\phi,$ equation
(\ref{bd.99}) becomes,%
\begin{equation}
\frac{d^{2}u}{dv^{2}}-3\left(  \frac{du}{dv}\right)  ^{2}+\frac{1}{v}\frac
{du}{dv}+\left(  3vV_{0}-2\right)  u^{3}=0,
\end{equation}
with analytic solution
\begin{equation}
\frac{1}{9}u^{2}\left(  v\right)  =\left(  6V_{0}v^{3}+9v^{2}-u_{1}+u_{2}\ln
v\right)  ^{-1},
\end{equation}
that is
\begin{equation}
\frac{1}{9}\dot{\phi}^{2}=\left(  6V_{0}\phi^{3}+9\phi^{2}-u_{1}+u_{2}\ln
\phi\right)  ^{-1}.
\end{equation}

Remark that solution (\ref{bd.100}) is an analytic solution for the
cosmological model since it has six integration constants, they are $\psi
_{0},~a_{0},~\Sigma_{1}$,~$\varpi_{0},~\varpi_{1}$ and $t_{0}$, while the
essential integration constants are~$\varpi_{0},~\varpi_{1}$. In a similar
way, we can construct more analytic solutions for the two cosmological models
of our consideration

This work contribute to the subject of integrability for modified theories of
gravity. Such an analysis is important in order to study and determine the
existence of some important solutions which describe different physical
conditions in the cosmological evolution.

Furthermore, the results of this analysis can be used to prove the existence
of actual solution for the field equations. which is an important result in
order to be able to understand the physical properties for a given
cosmological model, as we shall know either the numerical results we get from
a model are not sensitive on small changes of the initial conditions and a
chaotic behaviour does not exist.

\end{document}